# Power law decay and phase rigidity for large-amplitude coherent phonons in bismuth at helium temperature: Possible evidence for transient supersolid state

O.V. Misochko, and M.V. Lebedev

*Institute of Solid State Physics, Russian Academy of Science*
*142432 Chernogolovka, Moscow region, Russia*
(Submitted April 8, 2010)

Intense ultrafast laser excitation can produce transient states of condensed matter that would otherwise be inaccessible. At high excitation level, the interatomic forces can be altered resulting in an unusual lattice dynamics. Here we report the study of coherent lattice dynamics in Bi made for various excitation levels at helium temperature. We demonstrate that under certain conditions the fully symmetric phonons of large amplitude exhibit anomalous decay and phase rigidity, both of which possibly signaling the attainment of transient supersolid state.

## I. INTRODUCTION

One of the most mysterious phases of matter is the supersolid state where the diagonal long-range order responsible for the ordering in real space, coexists with the off-diagonal long-range order associated with the ordering in momentum space. Even though the question of the existence of superfluidity in a crystal (or whether does solidity preclude superfluidity) came under theoretical scrutiny long ago [1], first rather positive experimental evidence for the supersolidity of helium was obtained only in recent years [2]. When the supersolidity appears as a result of the crystallization of a quantum liquid due to the interplay between repulsive and attractive interatomic forces, the supersolid state is called a coherent crystal [3]. The necessary condition for coherent crystal realization is provided by bounded interactions with a partly negative Fourier transform. It is the negative minimum in the interatomic potential that is responsible for the fact that in the coherent crystals the lattice density may not adjust to that of the atoms but follow it with some delay in such a way that both densities support each other [3]. To fully appreciate the peculiarities of a coherent crystal it is enough to compare it with an "ordinary" one. The ordinary crystal consists of localized identical atoms making small oscillations about their respective equilibrium position. Thus, each atom is tagged by a lattice site, which means a built-in violation of permutation invariance. If we remove the latter restriction, we obtain the crystal where the atoms can move freely. The defining characteristic of a coherent crystal is that its lattice constant being not determined by the mass density but by the global properties of the inter-atomic interaction.

The coherence responsible for any of the macroscopic quantum phenomena such as Bose-Einstein condensation, superfluidity, superconductivity, and lasing can be spontaneous or driven by an external source. Only in the latter case, to which most likely coherent phonons belong to [4], the coherence understood as macroscopic occupation of a single quantum state occurs far from equilibrium. It is recent works on condensation of different short-lived quasiparticles in solids that have opened up the new field of non-equilibrium quasiparticle condensates [5]. As far as lattice excitations are concerned, their coherence can be achieved irradiating a crystal by ultrashort laser pulses. Indeed, first-order coherence of a phonon ensemble created by ultrashort pulse has been qualitatively shown by observing an interference pattern when two independent phonon ensembles created by two separated in time pump pulses were brought to overlap [4,6].

## II. BISMUTH PROPERTIES

Among different crystals, coherent phonons have been most extensively studied for bismuth, both experimentally and theoretically [4]. Bismuth structure, responsible for its semimetalic behavior, is usually derived from a cubic lattice rhombohedrically distorted along the body diagonal by the Peierls-like mechanism [7]. If the coincidence of the Fermi surface and the zone boundary were exact, one band would be completely filled and the next higher one completely empty, making the crystal an insulator. Actually in bismuth, a few electrons spill over into the higher band (at L - point of the Brillouin zone), leaving a few holes in the lower one (at T - point). As these electrons and holes are in regions of very great curvature of the energy surface, they have low effective mass [7]. The order parameter for low symmetry state is the internal (Peierls) shift $\delta$, or what is equivalent – the inverse gap (the energy separation between hole and electron bands). Indeed, the carrier density in Bi is well correlated with the internal shift: the larger is the shift, the larger is the density. This is also related to the magnitude of the band overlap – the larger shift results in the larger volume of the carrier pockets. The fully symmetric $A_{1g}$ phonons of this state correspond to the out-of-phase oscillation of the two atoms in the unit cell against each other along trigonal axis, and thus modulate the order parameter of semimetal phase.

## III. EXPERIMENTAL DETAILS

The sample used in this study was a single crystal of bismuth with surface containing the trigonal axis. The crystal was mounted into a cryostat with a silver paste and all measurements were made at *T=5K*. A *Ti*:sapphire mode-locked laser oscillator at 800 nm was amplified using regenerative *250 KHz* amplifier pumped by frequency-doubled *Nd:YVO₄* laser. The amplified and compressed laser pulse had the duration of *45 fs* at sample position. The laser beam was divided into pump and probe parts polarized perpendicular to each other. Both the pump and probe beams were kept close to normal incidence and focused to a spot with diameters of *0.08* and *0.04 mm*, respectively, by a single 5 *cm* lens. The excitation geometry $\vec{E} \parallel c$ precludes the excitation of doubly degenerate $E_g$ phonons and thus allows isolating the fully symmetric coherent dynamics. To carry out coherent control experiments, we used a pulse shaper, which provides the modulation of the amplitude and phase of the pump pulse spectrum. A more detailed description of the experiments can be found in [6, 8].

## IV. EXPERIMENTAL RESULTS

Before describing our experiments it is perhaps useful to present a simple picture of what occurs in a typical ultrafast pump-probe experiment. We excite the crystal creating a lot of phonons and electrons in a time short compared to phonon lifetimes and their inverse frequencies. The first condition means that we are dealing with transient lattice state, while the second is responsible for coherent nature of the lattice excitation (the first excited and vacuum lattice states are in a coherent superposition). Since in rhombohedrical Bi the internal shift $\delta$ has the same symmetry as the $A_{1g}$ phonon mode, the excitation of the latter results in the order parameter modulation $|\delta - A(t)|$, where $A(t)$ is the modulation amplitude. Thus, increasing the excitation we move towards the cubic phase. As the internal shift in Bi is quite small $\delta = 0.19\,\overset{o}{A}$, even below the Lindemann stability limit we can transiently restore the spontaneously broken cubic symmetry. However, after the excitation, the crystal inevitably relaxes back to equilibrium (rhombohedrical) state that is $|\delta - A(t)|_{t\to\infty} \to 0.19\,\overset{o}{A}$ as $A(t) \to 0$. Therefore, observing how the order parameter evolves in time, we can study transient states on the way to equilibrium.

Let us investigate the pump dependence first. We present the results in the form of the dependence of the coherent amplitude on the pump fluence, which was varied in our study by almost four orders of magnitude, see Fig. 1. If the linear dependence (observed at low pump strength) is extrapolated to the high strength region, the observed dependence $A = f(F)$ is naturally divided into three components: linear, superlinear, and sublinear, each exhibiting its characteristic dynamics. Below, we will primarily focus on the discussion and comparison of the linear and nonlinear parts.

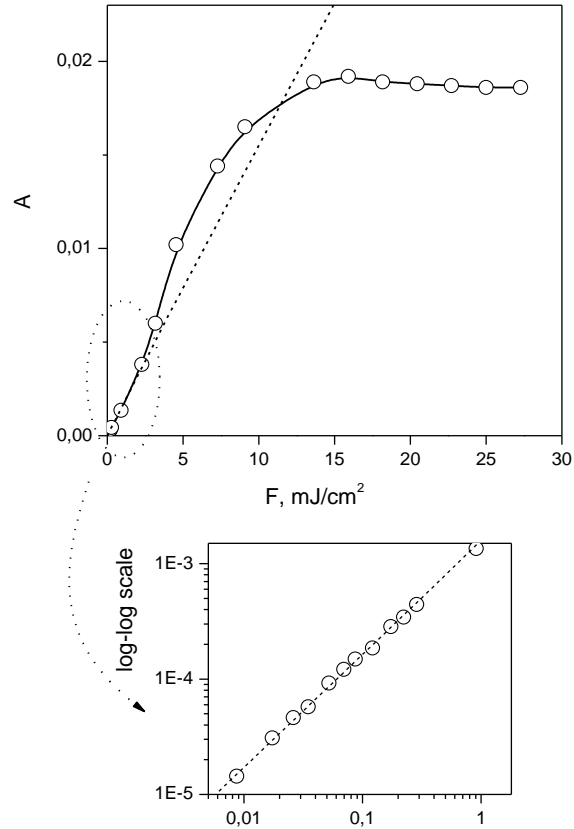

FIG.1: Coherent amplitude versus the energy density of the pump pulse. The dashed line is the extrapolation of the linear regime shown in the lower panel in the log–log scale to the region of high intensities.

Now let us compare the lattice relaxation for low- and large-amplitude coherent phonons of $A_{1g}$ symmetry. Before the excitation the lattice is in the ground state as our temperature is quite low ($T=5K$). After excitation, due to a broad spectrum of the pump pulse mixing the ground and first excited state, we place the lattice into a coherent state decaying in time. At low amplitude this decay is exponential, therefore a decay time $\tau$ and a decay rate $k = \tau^{-1}$ can be easily defined [8]. The both are independent of excitation level and agree well with those observed in spontaneous Raman scattering [4]. At moderate excitation levels the decay remains exponential, but with the characteristic decay time turning into a function of excitation. The reduction in the lifetime of the coherent phonons can be easily explained: when the large-amplitude coherent phonons decay, they produce highly excited acoustic modes in certain regions of the Brillouin zone, resulting in an increase in the decay into these modes. At high levels of excitation, the observed decay is substantially modified, compared to low-amplitude motion. As can be seen from Fig. 2, the relaxation for large amplitude coherent phonons follows a fast decay at short times but it is slower than an exponential at long times. The log-log plot for large-amplitude coherent phonons reveals a straight line at long times characteristic of a long time tail, whereas that for small-amplitude coherent phonons exhibits noticeable curvature over the whole range of time scales. Thus, with increasing coherent amplitude the relaxation changes from an exponential to a power law. The latter law means that for the strongly excited lattice there is no typical temporal (or spatial) scale in the conventional sense, or that certain derivatives of thermodynamical potentials diverge. Thus, such a power law relaxation is indicative of a phase transition, similar to the case of pumped excitons [8].

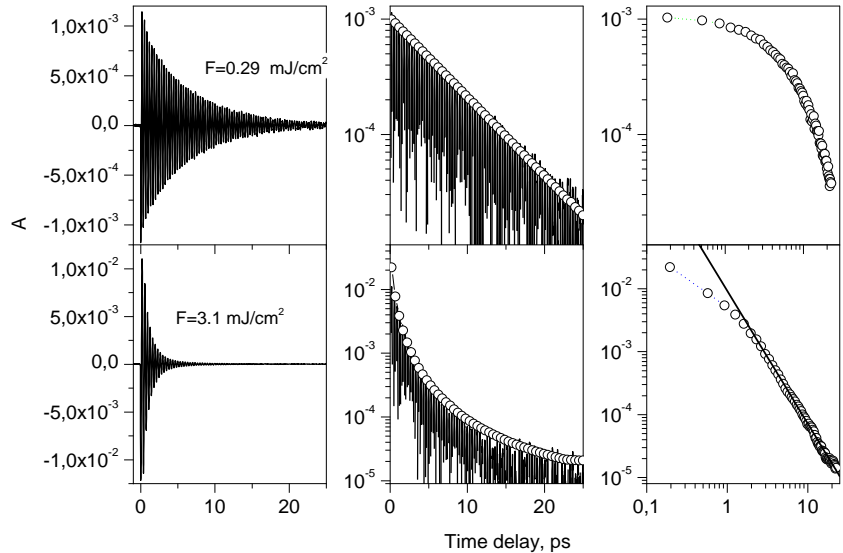

FIG.2: Coherent $A_{1g}$ oscillations at helium temperature for low (upper panels) and high (low panels) excitation levels. The right panels show the data in a semi-logarithmic and logarithmic scale.

Next we will address phase properties of the coherent oscillations. For small amplitude coherent phonons, the choice of the delay time between two successive pump pulses makes it possible to enhance or completely cancel oscillations [6]. Such a process can be represented as the sum of two interfering ensembles of coherent phonons created at different times and based on the model of displacive excitation of coherent phonons [9] (DECP) understood as follows. The first pump pulse results in a change in the equilibrium interatomic separations. More exactly, photoexcitation changes the potential on which atoms move keeping the atoms immobile, since electromagnetic field does not couple directly to lattice due to a huge energy mismatch between photons and phonons. As a result, the atoms begin to move to a new, displaced equilibrium and, owing to their inertia, continue to be in motion after reaching it, leading to oscillations. To cancel the oscillations, one should wait until the atoms are at the opposite slope of the potential at a turning point and, then exciting a necessary number of carriers to higher bands, shift the potential minimum to this point. Since the kinetic energy is zero at the turning point, the atoms being at the potential minimum cease their movement. For the maximum enhancement, the potential should be shifted at the times that are multiples of the oscillation period. The shift at the intermediate times leads to oscillations with an amplitude smaller than the maximum one.

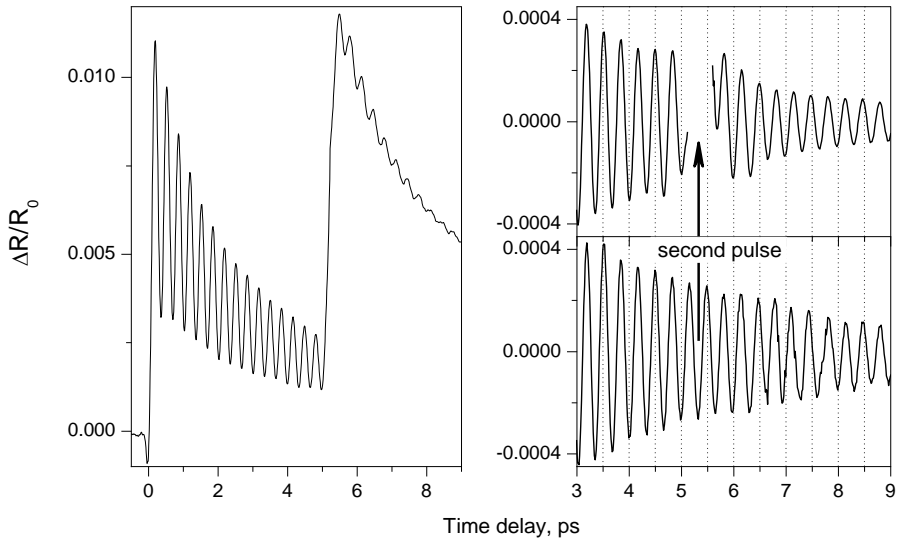

FIG.3: Time-resolved reflection of the Bi single crystal at T=5 K for strong double pulse excitation. In this case, the excitation level was taken in the superlinear region, where the oscillation amplitude increases faster than the excitation energy density, but below the threshold for collapse–revival effect. Note that the second pump pulse almost does not change the oscillation amplitude, although the energy of the second pulse coincides with that of the first one. The panels on the right show the oscillations for double (upper panel) and single pulse (bottom panel) excitation; the comparison reveals that the oscillation phase remains unchanged after the action of the second pump pulse the arrival of which is marked by an arrow.

Quite unexpectedly, this simple and well-understood picture fails completely for large amplitude motion provided the motion is excited by ultrafast pulses that are polarized along the trigonal axis. Figure 3 illustrates the double pulse excitation for large amplitude with a control parameter (interpulse separation) *γ = 5.31 ps*. One can easily see that the second pump pulse leads to a significant change in the non-oscillatory (background) component, but the oscillation amplitude varies only slightly, although the second pulse at the given control parameter interacts with the crystal at the time when atoms are at the classical turning point for which interference is constructive

It is even more surprising that phase of the large-amplitude coherent phonons in this case remains essentially unchanged after the second pump pulse. This is illustrated in Fig. 3(b), where the oscillatory parts for the single and double pulse excitation are shown. Their comparison reveals that the phase of the double pulse signal after the second pulse coincides with that of the single pulse one. Varying the interpulse separation $\gamma$ within $\pm T/2$, where $T$ is the phonon period, we found that the phase rigidity is observed for any time instant of the interaction of the second pulse with the crystal, *i.e.,* for any of the slopes of the potential at which atoms are located. However, the phase rigidity cannot be realized for the light polarizations lying within the basal plane $\vec{E} \perp c$ [10]. Moreover, even for $\vec{E} \parallel c$, the phase rigidity appears not instantaneously with large amplitude motion, but with some delay: coherent control for several first oscillations reveals that only the linearity of interference is violated for large amplitude motion. Interestingly, the time needed for the phase to become rigid approximately coincides with the time where the decay switches from exponential to power law, that is $\tau_{rigid} \approx 2 ps$.

Staying within the DECP model, the observed phase rigidity indicates that the second pump pulse shifts both the potential, on which atoms move, and the atoms themselves; i.e., atoms performing large amplitude motion become "glued" to the potential, the shape and location of which is determined primarily by the density of photo-induced carriers. Thus, the phase rigidity for large amplitude coherent phonons can be explained only by the fact that the motion of atoms during the action of the second pulse is adiabatic; *i.e.*, the state of the bismuth lattice after the second pump pulse remains essentially unchanged.

The persistence of oscillations after the second pulse shows that the atoms undergo a finite motion with respect to a stable quasi-equilibrium position as required by the balance between the restoring force and their inertia. The impossibility of modifying the phase of this motion could be interpreted as an abrupt increase in the inertia (mass of the equivalent oscillator), but this interpretation will be correct only if we increase simultaneously the rigidity of the oscillator by the exactly same amount to keep the oscillation frequency unchanged. Since the DECP mechanism is kinematical (the action on the oscillator is performed through the displacement of the suspension point of the spring or pendulum filament), we face a paradoxical situation. The phase rigidity suggests (taking into account the energy conservation) that the second pulse initiates the motion of the atoms with respect to a *neutral* equilibrium. However, this is possible only in the case of absolute rigidity of the system, which, as we suggest, can appear due to the synchronization of the collective motion of the charge carriers (plasmons) and atoms (phonons).

V.  **DISCUSSION**

The synchronization may be illustrated by means of a toy model, which, while it would be quite difficult to treat rigorously, is nevertheless quite a useful guide to the behaviour in more realistic cases. In particular, it illustrates rather straightforwardly the fact that the collective motion of mobile carriers and atoms in bismuth can be synchronized. Since bismuth has free carriers, *i.e.* there is free Fermi surface; there are two distinct kinds of state of motion. In the first case, we can have uniform motion of the whole crystal, which requires dynamical excitation. In the second case, when the excitation is kinematical, we allow the crystal lattice to be at rest and just move the Fermi surface, which is equivalent to taking atoms from one side of it and replacing them on the other. The shifted Fermi surface is equivalent to exciting a charge-density wave, and if we manage to synchronize the movement of the charge-density wave and that of the atoms, their mutual sliding behaviour may result in anomalous dynamics observed for large amplitude motion in bismuth.

Thus, our task is to couple amplitude mode of charge-density wave to phonon mode. However, the plasma frequency $v_{pl}$ in unexcited crystal is around *5 THz* and photoexcited carriers increasing the carrier density n are expected to only raise it as $v_{pl} \propto en^{1/2}m^{-1/2}$, where *e* is the charge and *m* is the effective mass. To synchronize the electronic and atomic motions one must *decrease* the plasma frequency bringing it to match that of collective atomic coordinate (*3 THz*). This can be in principle achieved because coherent phonons in bismuth not only modulate the gap between rigid bands (deformation potential, interband transitions), but also change the slope of the band dispersion around the Fermi surface due to Pomeranchuk effect [11] caused by intraband transitions, see Fig.4. In this case, the larger mass of the carriers after the excitation can bring the electronic and lattice frequency to resonance since electronic system in the vicinity of the Pomeranchuk instability may have peculiar properties due to a "soft" Fermi surface, which can be easily deformed by anisotropic perturbations.

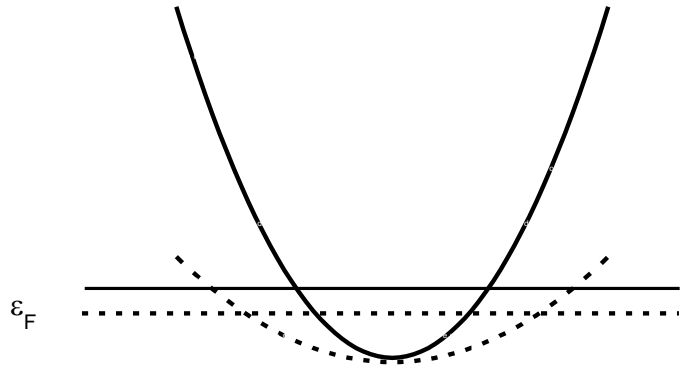

FIG.4: Pomeranchuk instability: modulation of the slope of the band dispersion around the Fermi level. Solid lines –unexcited crystal, dotted lines – excited crystal.

Given that the charge-density wave in bismuth can be incommensurate, we can realize the situation in which in the lattice constant and the density can change independently. Indeed, in our experiments we have a fixed atomic density; however, we allow the lattice constant and thus the incommensurability to vary, driven by the dependence of the Fermi surface on the incommensurability. As a result, we are dealing with unusual pairing: for $\max(|\delta - A(t)|)$ the atoms are paired in real space, whereas for $|\delta - A(t)| \to 0$ the pairing takes place in reciprocal (momentum) space, that is the phonons running in opposite directions along the trigonal axis are Bose-condensed [6]. The latter is equivalent to pairing in the particle-antiparticle channel described by Kirzhnits and Nepomnyashchii [2]. The whole situation with the alternating real-reciprocal space pairing is kind of the quantum analogue of the classical Kapitza pendulum [13], where intense vibrations of the suspension point results in additional, dynamically stable equilibrium for the inverted (upside–down) position. Then, the superposition of the straight–down and upside–down positions (that is static and dynamically equilibrium states) can be consider as a diagonal order, while the superposition of two unstable states, which appear simultaneously with the dynamically stable one, can be treated as off-diagonal order. The condition for the onset of dynamical stability in the case of Kapitza's pendulum:

$$2\pi^2 a^2 \nu^2 \geq gL(1 + \frac{K^2}{L^2}); \quad a << L; \quad \nu >> \frac{1}{2\pi}\sqrt{\frac{g}{L}} \qquad (1)$$

where $a$ is the amplitude and $\nu$ is the frequency of the pivot vibrations, $g$ - gravitational constant, $L$ - pendulum length, and $K$ - inertial length of the pendulum. This inequality just reflects the fact that the kinetic energy for vibrating suspension point is larger than the potential energy of upward pointing pendulum. The conditions (1) should coincide with the onset for coherent crystallization, which for a Bose-system reads [2]:

$$\gamma_{k_0} = -\frac{4mn\nu(k_0)}{k_o^2} - 1 > 0 \qquad (2)$$

where m is the atomic mass, n is the atomic density, $\nu(k_0)$ is the negative Fourier component of the interatomic potential, and $k_0$ is the wave vector for which the attraction is maximal. The inequality (2) means that the kinetic energy of a pair is larger than the interaction energy even though the total interaction energy per pair can be larger than the kinetic energy - $m|\nu(k_0)|k_0 << 1$. The latter condition is satisfied since the system is compressed and there are many atoms within the interaction volume.

## VI. CONCLUSIONS

To summarize, we have observed that while the coherent phonon amplitude vanishes exponentially for very small excitation, it relaxes toward a quasi-stationary value, when the excited phonon density exceeds some critical value. But in contrast to the subcritical behaviour, the approach to the equilibrium is not exponential anymore and well described by a simple power law. The phase properties of large amplitude coherent phonons are also unexpected: after some time their phase becomes rigid and cannot be changed by a second pump pulse. The phase rigidity and power law decay are possible indications of broken symmetry, seemingly signaling the attainment of a transient supersolid state. The microscopic origin of these effects is not completely understood and we hope that our results may stimulate corresponding theoretical investigation.

**ACKNOWLEDGEMENT**

This work was supported by the Russian Foundation for Basic Research (project no. 10-02-00506).

**References**

* Electronic address: misochko@issp.ac.ru
[1] A.F. Andreev, and I.M. Lifshitz, Sov. Phys. JETP **29**, 1107 (1969).
[2] D.A. Kirzhnits, and Yu.A. Nepomnyashchii, Sov. Phys. JETP **32**, 1191 (1971).
[3] E. Kim, and M.H.W. Chan, Nature **427**, 225 (2004).
[4] K. Ishioka, and O.V. Misochko, In Progress in Ultrafast Intense Laser Science V, Eds. K. Yamanouchi, A. Giullietti, and K. Ledingham, 23-64, Springer Series in Chemical Physics, Berlin (2010).
[5] D. Snoke, Nature **443**, 403 (2006).
[6] O.V. Misochko, and M.V. Lebedev, JETP Letters **90**, 284 (2009).
[7] R. E. Peierls, More surprises in theoretical physics (Princeton Univ. Press, 1991).
[8] O.V. Misochko, and M.V. Lebedev, JETP **109**, 805 (2009).
[9] O.M. Schmitt, *et. al.*, Eur. Phys. J. **B16**, 217 (2000).
[10] H.J. Zeiger, *et. al.*, Phys. Rev. **B 45**, 768 (1992).
[11] K. Ishioka *et. al.*, unpublished results
[12] I.Ia. Pomeranchuk, Sov.Phys. JETP **35**, 524 (1958).
[13] D. ter Haar (ed) Collected Papers of P.L. Kapitza, 94-101, Pergamon, Oxford (1965).